\documentclass[twocolumn,prc,floatfix,showpacs,preprintnumbers,amsmath,amssymb,merge]{revtex4}

\usepackage{graphicx}
\usepackage{dcolumn}
\usepackage{bm}
\usepackage{natbib}

\def \lf {\left}
\def \ri {\right}

\begin{document}



\title{Influence of the single-particle structure on the nuclear
       surface and the neutron skin
}

\author{M. Warda\textsuperscript{1}}
 \email{warda@kft.umcs.lublin.pl}
\author{M. Centelles\textsuperscript{2}}
 \email{mariocentelles@ub.edu}
\author{X. Vi\~nas\textsuperscript{2}}
 \email{xavier@ecm.ub.edu}
\author{X. Roca-Maza\textsuperscript{3}}
 \email{xavier.roca.maza@mi.infn.it}
 
\affiliation{
\textsuperscript{1}Katedra Fizyki Teoretycznej, 
Uniwersytet Marii Curie--Sk\l odowskiej,
ul. Radziszewskiego 10, 20-031 Lublin, Poland\\
\textsuperscript{2}
Departament d'Estructura i Constituents de la Mat\`eria
and Institut de Ci\`encies del Cosmos (ICC),
Facultat de F\'{\i}sica, Universitat de Barcelona,
Diagonal {\sl 645}, {\sl E-08028} Barcelona, Spain\\
\textsuperscript{3}
Dipartimento di Fisica, Universit\`a degli Studi di Milano and INFN, Sezione di
Milano, via Celoria 16, I-20133 Milano, Italy
}

\date{\today}

\begin{abstract}
We analyze the influence of the single-particle structure on the
neutron density distribution and the neutron skin in Ca, Ni, Zr, Sn, and Pb
isotopes. The nucleon density distributions are calculated in the
Hartree-Fock+BCS approach with the SLy4 Skyrme force. A
close correlation is found between the quantum numbers of the valence
neutrons and the changes in the position and the diffuseness of the nuclear
surface, which in turn affect the neutron skin thickness. Neutrons in the
valence orbitals with low principal quantum number and high angular momentum
mainly displace the position of the neutron surface outwards, while neutrons with high
principal quantum number and low angular momentum basically increase the
diffuseness of the neutron surface. The impact of the valence shell neutrons on
the tail of the neutron density distribution is discussed.
\end{abstract}

\maketitle


\section{Introduction}


The spatial distribution of nucleons inside a nucleus is one of the most basic
topics in nuclear physics. The density distribution of protons is quite well
mapped from the numerous experiments of elastic electron-nucleus and
muon-nucleus scattering \cite{fri95,vri87} and the charge radii of many
nuclei are known with uncertainties well below~1\% \cite{angeli13}.
Neutrons, as neutral particles, are much harder to resolve and knowledge
about their spatial layout in a nucleus is still limited. Until now, neutron
radii have been measured in less than thirty isotopes
and the experimental neutron density distribution is known only in a
few nuclei with relatively large error bars
\cite{hof80,cla03,bat95,kra04,trz01,jas04,klo07,kli09,zen10}.
Thus, information about the layout of neutrons inside a nucleus
often comes only from the theoretical predictions at present.
However, the new experimental advances in techniques such as
elastic proton scattering \cite{zen10} and coherent
pion photoproduction from nuclei \cite{wat12,tar13}, and the advent of
parity-violating elastic electron scattering facilities
\cite{abr11,horowitz14,mesa}, suggest that largely improved determinations
of neutron radii and neutron density distributions may be
possible in the near future.

Knowledge of neutron distributions is very important as it constitutes a
necessary input in a wide range of problems in physics. It is strongly related
to the isospin properties of nuclear forces and the nuclear symmetry
energy \cite{cen09,war09,che05,fur02,kor13,zhang13}.
%
%
The profile of the neutron density distribution is demanded as an input in the
analysis of many scattering experiments. The arrangement of neutrons in nuclei
is important for collective nuclear excitations \cite{piekarewicz14,colo14}, such as the giant
dipole resonance \cite{tri08,roc13} and  pygmy dipole resonance
\cite{kli07,roc12,vre12}. Precise knowledge of the neutron skin thickness
(NST), i.e., the difference between the neutron and proton root mean
square (rms) radii:
\begin{equation}
\label{skin}
\Delta r_{np}= \langle r^2 \rangle_n^{1/2}
             - \langle r^2 \rangle_p^{1/2} \;,
\end{equation}
is not only of interest in nuclear structure physics. This quantity is strongly
correlated, within the realm of nuclear mean-field theories
\cite{bro00,typ01,die03,prov07,cen09,cen10,roc11,vinas14,gai12}, with the slope of the
nuclear symmetry energy at saturation density and therefore may be used to
constrain the equation of state of neutron-rich matter. Thus, the
results of the investigation of the distribution of neutrons in atomic nuclei
affect studies of such distant areas of physics as heavy-ion collisions
\cite{gai04,she07,li08,tsa09}, scattering of polarized electrons on nuclei
\cite{vre00,hor01a,roc11,abr11,moy10,liu12,fin13}, precision tests of the
standard model by atomic parity violation \cite{vre00apv,sil05}, and nuclear
astrophysics~\cite{hor01,lat04,lat07,prov11,khoa11}.

Theoretical predictions of neutron density distributions can be verified, in
principle, by the comparison of the calculated values of the neutron 
rms radii and of the NST with the available experimental data.
The uncertainties in the measured values leave much freedom for theoretical
neutron density distributions \cite{war98,jas04,trz01,jia07}. Moreover,
rms radii and the NST are general properties of the neutron density
distribution; different density profiles (e.g. those calculated with various nuclear
forces) may give the same values of the neutron rms radius and the NST
\cite{war10}. Therefore, careful theoretical studies are required to  understand
the physics of nucleon distributions inside a nucleus. 
The main issues affecting the deviations of the NST from the picture of a 
smooth variation with the neutron excess of the nucleus are nuclear
deformation \cite{war98} and the quantum-mechanical properties of the 
nucleonic orbitals. In the present article we concentrate on the latter 
effect.

The basic features of the neutron skin of nuclei can be explained by the nuclear 
droplet model (DM) \cite{mye80,swi05}. The DM predicts that the NST grows
on average linearly with the relative neutron excess $I=(N-Z)/A$, which was
confirmed by the experimental data of Refs.\ \cite{trz01,jas04}. In the standard
version of the DM the neutron and proton surface diffusenesses are assumed to be
equal, although the influence of different surface diffusenesses between
neutrons and protons was also investigated \cite{swi05,war09,vin12}. In
particular, it has been shown that the surface contribution to the NST in
nuclear mean-field models is not negligible
\cite{war09,war10,cen10,vin12,miz00};
indeed, self-consistent mean-field calculations predict in many neutron-rich
isotopes twice thicker neutron than proton surface diffuseness. Moreover, abrupt
changes of the surface diffuseness between isotopes may appear (see Figs. 7 and
8 of Ref.\ \cite{war10}). It has been found that the surface of the neutron
distribution is narrower in the doubly magic nuclei, whereas it extends over a
wider region in the mid-shell nuclei. Such a behavior suggests that quantum
shell effects impinge on the NST on top of the macroscopic DM predictions.

There is significant interest currently in the exploration of the properties
of exotic nuclei as the radioactive ion beam facilities in laboratories
worldwide are extending the nuclear landscape to new limits.
We have devoted some effort in previous works \cite{scattering08,scattering13}
to study the evolution of the nuclear charge density from stable to
exotic nuclei, and its relation with the changes of the proton
shell structure in isotopic and isotonic chains.
In the present article we investigate the influence of the single-particle (sp)
properties of the valence neutrons on the nuclear surface and the
neutron skin of stable and unstable nuclei. We study
several isotopic chains representative of different mass regions in order to
examine the changes of the neutron skin when subsequent neutrons are added into
particular orbitals. We analyze two mechanisms of generating the neutron
skin. One of them arises from a displacement between the positions of the
equivalent neutron and proton sharp surfaces (it is mostly a ``bulk'' effect).
The other one is a consequence of different surface diffuseness between the
neutron and proton density profiles (it is mostly a ``surface'' effect). To
minimize the influence of deformation
on the results \cite{war98} it is helpful to choose nuclei with magic proton
number, which are mostly spherical nuclei in their ground states. Hence, we
take into consideration the Ca, Ni, Zr, Sn, and Pb elements. We focus our study
mainly on the neutron-rich nuclei because
the neutron skin is larger and more sensitive to isotopic effects there. 
We concentrate on the last full major shell (or shells) of the considered elements.
In this way we examine the neutron skin in Ca and Ni from $N=20$ to $N=50$, in
Zr and Sn from $N=50$ to $N=82$, in Sn between $N=82$ and $N=126$, and in Pb from
$N=126$ to $N=184$. We start with the analysis of the Sn isotopes ranging 
from $N=82$ to 126 because the
discussed sp properties are quite well magnified and easy to describe in this
relatively long chain. After this illustrative example  we study the other
elements.

The structure of this article is the following. In Sec. II the basic ideas of our
theoretical description of the NST are collected (an extended presentation can
be found in Refs.\  \cite{war10,cen10}). The detailed analysis of the results
for the NST in the Sn isotopes is presented in Sec. III. The other isotopic
chains are discussed in Sec. IV. Finally, the conclusions are presented in Sec.
V.


\section{Bulk and surface contributions to the neutron skin thickness}

In our study we compute the density distributions of nucleons in even-even
isotopes in self-consistent Hartree-Fock calculations using the SLy4 nuclear
functional \citep{cha98,*cha98a}. SLy4 is a Skryme-type force that was developed with a
focus on neutron-rich nuclei and the equation of state of neutron matter.
It has been successfully applied in studies of a wide range of
phenomena and it is known to describe reasonably well nuclear properties such as
masses, deformations, nucleon separation energies, and radii along the periodic
table \cite{erl12}.
Though the general features of the NST described in this paper are basically
independent of the nuclear mean-field interaction used to compute them,
the fine details can depend to a certain extent on the nuclear interaction.
In our calculations we assume spherical symmetry for all considered
nuclei (however, some of the Zr isotopes are known to be deformed in
their ground state). We use a volume BCS pairing with pairing strengths
$V_0^n=300$ MeV and $V_0^p=360$ MeV. 
The pairing window is taken to be $1\;\hbar\omega$ above the Fermi level.
To fit the pairing strengths we used as reference data the SLy4 results of the
HFB+Lipkin-Nogami model of Ref.~\cite{sto03} for the binding energies
of the mid-shell nuclei $^{116}$Sn for neutron pairing and $^{144}$Sm for proton pairing.
The treatment of the continuum for neutron-rich nuclei close to the drip line
is done following the prescription given in \cite{est01}. In this way the HFB
energies reported in Ref.~\cite{sto03} are overall well reproduced from the proton
to the neutron drip line. We find the two-neutron drip line of the elements
considered in the present work at $^{68}$Ca, $^{78}$Ni, $^{122}$Zr, $^{176}$Sn, and
$^{266}$Pb, which is the same result as in Ref.~\cite{sto03} excepting that
the drip line nucleus $^{174}$Sn of \cite{sto03} is shifted to $^{176}$Sn in
our calculation.

To get a better grasp of the properties of the neutron skin it can be useful to fit
the density profiles obtained in the self-consistent mean-field calculations
by two-parameter Fermi (2pF) distributions \cite{vri87,has88}:
\begin{equation} 
\label{2pf}
\rho(r) = 
\frac{\rho_0}{1+ \exp{ [(r-C)/a] }} ,
\end{equation} 
where $\rho_0$ is the central density, $C$ is the half-density radius, and  $a$
describes the surface diffuseness. As a result, one obtains numerically the two
most important quantities characterizing the shape of the density profiles,
namely the position and the thickness of the nuclear surface. Both of them are
crucial for a proper determination of the NST, as the NST is defined through the rms
radii and consequently it is very sensitive to the density profile at the
surface. There is no unique prescription to parametrize a given density profile
with a 2pF function. Following earlier works
\cite{war10,cen10}, we fit the $a$, $C$, and $\rho_0$ parameters to reproduce
the quadratic and quartic moments of the neutron or proton density distribution,
and the number of nucleons. It has been shown that this
method reproduces with good accuracy the surface region of any realistic density
profile given as an input \cite{war10,cen10}.

The neutron skin can be easily understood assuming 2pF distributions for both
neutrons and protons. 
Indeed, Fermi-type densities are common in the 
extraction of neutron skins from different experiments, as in the case 
of neutron skins deduced from exotic atoms \cite{trz01,fri05,fri09} 
or from coherent pion 
photoproduction cross sections \cite{wat12,tar13}. Within the context of 2pF 
densities, it has become popular to discern two main scenarios for the 
neutron skin of nuclei.
In the first scenario, the neutron skin is formed when the
neutron half-density radius is larger than the proton half-density radius and the surface
diffusenesses of neutrons and protons are the same (i.e., $C_n>C_p$ and $a_n=a_p$).
Such 2pF density profiles are called a ``skin" type distribution \cite{trz01}. The alternative
scenario assumes that the differences between both density profiles are due to
an enlarged neutron surface diffuseness with the same neutron and proton
half-density radii (i.e., $a_n>a_p$ and $C_n=C_p$). The corresponding 2pF
density profiles are called a ``halo" type distribution \cite{trz01}. In general,
both situations may coexist in a nucleus, simultaneously contributing to the NST
\cite{swi05,cen09,war09,war10,cen10}.

Despite the simplicity of describing the density profiles through 2pF
distributions, we have shown in previous papers \cite{war10,cen10} that the
half-density radii $C_n$ and $C_p$ are not the most appropriate radii for
extracting the bulk and surface contributions to the NST. Following Refs.\
\cite{war10,cen10} we introduce the bulk contribution as
\begin{equation}
\label{rb1}
\Delta r_{np}^{\rm bulk} \equiv \sqrt{\frac{3}{5}}\left(R_n-R_p\right) \;,
\end{equation}
where $R_n$ and $R_p$ are the neutron and proton equivalent sharp radii,
respectively \cite{has88}. The equivalent sharp radius $R$ corresponds to a sharp distribution
with a uniform density, equal to the bulk value of the actual density, having
the same number of particles \cite{has88}. As it can be seen in Fig.~1 of
Ref.~\cite{cen10} and Fig.~2 of Ref.~\cite{war10} (see also Ref.~\cite{has88}),
a sharp sphere with radius $C$ overestimates the original mean-field density in the
whole nuclear interior, whereas a sharp density distribution with radius $R$ is able
to reproduce properly the bulk part of the original density profile.
Therefore, the sharp radius $R$ rather than the
half-density radius $C$ is the suitable radius to describe the
size of the bulk region of the nucleus \cite{has88,war10,cen10}.

It is possible to express $R$ in terms of the
parameters $C$ and $a$ of the 2pF distributions \cite{has88,war10,cen10}, so that the
bulk contribution to the NST given by Eq.~(\ref{rb1}) can be written also as
\begin{equation}
\label{rb2}
\Delta r_{np}^{\rm bulk} \simeq  \sqrt{\frac{3}{5}}\left[(C_n -C_p)
+\frac{\pi^2}{3}
\left(\frac {a_n^2}{C_n}-\frac {a_p^2}{C_p}\right) \right] .
\end{equation}
The remaining part of the NST is the surface contribution:
\begin{equation}
\label{rs4}
\Delta r_{np}^{\rm surf} 
\simeq  \sqrt{\frac{3}{5}} \, \frac{5\pi^2}{6}
\left(\frac{a_n^2}{C_n}-\frac{a_p^2}{C_p}\right) .
\end{equation}
%
%
From Eqs.~(\ref{rb2}) and (\ref{rs4}), it is clear that the difference
between the half-density radii $C_n$ and $C_p$ can affect not only the
bulk contribution but also the surface contribution to the NST.
Similarly, the difference between the surface diffusenesses $a_n$ and
$a_p$ of the 2pF profiles affects $\Delta r_{np}^{\rm surf}$
and also $\Delta r_{np}^{\rm bulk}$. In general, both the changes of the
half-density radii and of the surface diffusenesses contribute simultaneously to
the NST.

Although the measurements provide data of the total NST only, from the
theoretical point of view it is interesting to analyze the separate
contributions of the bulk and surface parts of the NST. Useful information about the
nuclear surface can be obtained from such investigation. In the next sections we
analyze how the bulk and the surface contributions  change along the selected
isotopic chains and how they are correlated with the sp properties of the
valence neutrons.


\begin{figure}
\includegraphics[height=0.9\columnwidth, angle=270]{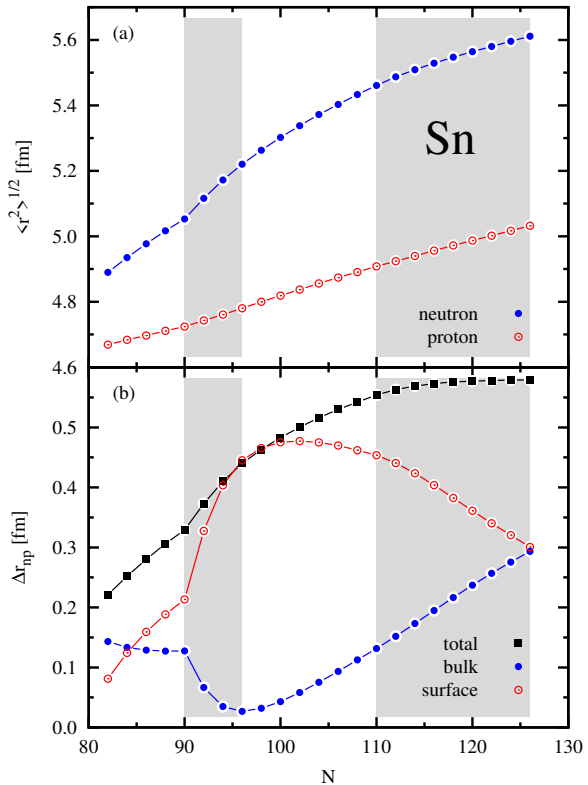}
\caption{\label{R_sn} (Color online) (a) Neutron and proton rms radii in Sn isotopes with
$82\le N\le 126$ calculated with the SLy4 Skyrme force. (b) The NST of
the Sn
isotopes (black squares), and the corresponding bulk (blue dots) and surface
(red circles) contributions to the NST [Eqs.\ (\ref{rb1}) and (\ref{rs4})].}
\end{figure}
\begin{figure}
\includegraphics[width=0.9\columnwidth]{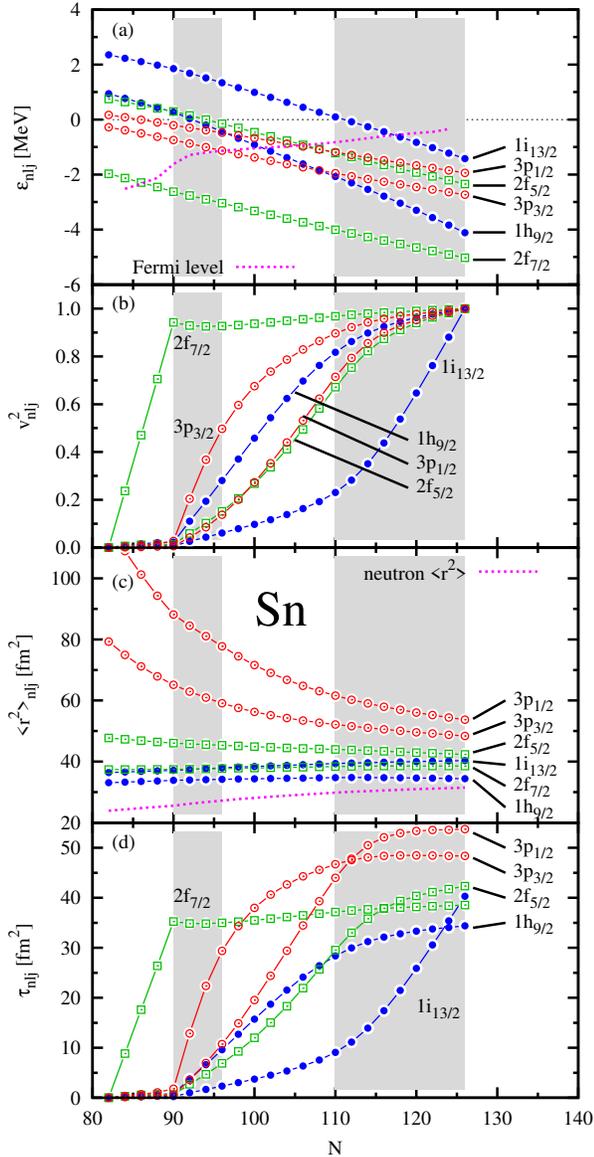}
\caption{\label{sp_sn} (Color online) Single-particle properties of the neutron orbitals 
belonging to the major valence shell for Sn isotopes with $82\le N\le 126$
calculated with the
SLy4 Skyrme force. (a) Single-particle energies and Fermi level, (b) occupancy
of each orbital, (c) mean square radii of each orbital defined in Eq.
(\ref{rnlj}) and total neutron mean square radius of the isotope, and (d)
contributions to the mean square radius of each level $\tau_{nlj}$ defined
in Eq. (\ref{tau1}). The $3p_{3/2}$ and $3p_{1/2}$ orbitals are marked  by
red circles,  the $2f_{7/2}$ and $2f_{5/2}$ orbitals by green squares, and the
$1h_{9/2}$ and $1i_{13/2}$ orbitals by blue dots.
}
\end{figure}

\begin{figure}
\includegraphics[width=0.9\columnwidth, angle=0]{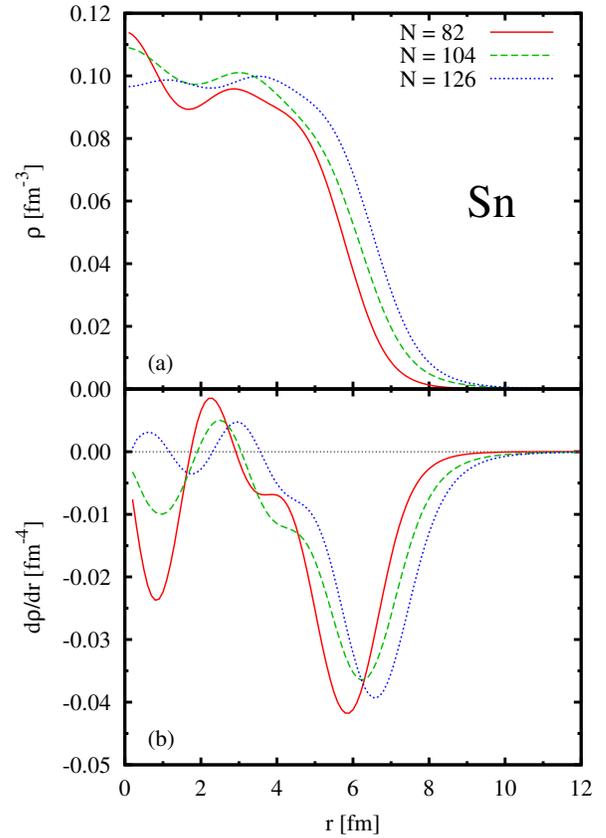}
\caption{\label{3dens} (Color online) Neutron density profiles of Sn isotopes with
$N=82$, $N=104$ and $N=126$ (a) and their derivatives (b).
\\[2mm] 
}
\end{figure}


\section{Neutron skin in the Sn isotopic chain for $82\le  N\le 126$}


The NST grows along the isotopic chain of a given element. The DM of Myers
and \'{S}wi\c{a}tecki \cite{mye6974,mye77,mye80} explains this behavior as
it predicts a faster linear increase with the relative neutron excess
$I=(N-Z)/A$ for the neutron rms radius than for the proton rms radius. The
rate of increase of the NST with $I$ is related to the density
dependence of the nuclear symmetry energy \cite{cen09,war09}.
Calculations with nuclear mean field models show that deviations from the
linear growth with $I$ can be found in the NST \cite{war10}. It is
easy to show that these deviations are connected with shell properties of
nuclei. Indeed, the NST displays local minima for magic neutron numbers,
whereas it exceeds the average trend for mid-shell isotopes (see Fig.~9 of
Ref.~\cite{war10}). Therefore, to explain this non-linearity, we have to
investigate the sp structure of nuclei. As a first example, we study
neutron-rich Sn isotopes of the major shell ranging from $N=82$ to $N=126$. 
The last isotope in this chain is the drip line nucleus.


\subsection{Correlation of neutron skin properties with quantum numbers of
valence neutrons}


In Fig.~\ref{R_sn}(a) it is easy to see that both the neutron and the proton rms
radii of Sn increase with increasing neutron number and that the slope for
neutrons is larger than for protons, as predicted by the DM.
The difference between the two curves of Fig.~\ref{R_sn}(a) is just the
NST plotted in panel (b) of this figure.
The proton radii show a rather linear dependence on the
neutron number, whereas some departure from linearity is observed in
the neutron radii. The same departure from linearity is therefore
observed in the NST of the Sn isotopes. Similar properties have been found in
Ref.~\cite{sch08} where an alternative splitting of the mean square
radius into geometrical and Helm radii is applied.

The bulk (\ref{rb1}) and the surface (\ref{rs4}) parts of the NST are also
plotted in  Fig. \ref{R_sn}(b). The lack of linearity visible in  the NST diagram
is magnified in the individual plots of these contributions. Four
intervals of
neutron number, marked in  Fig. \ref{R_sn} by white and grey stripes,  can be
easily distinguished in this plot. The first interval covers the region from
$N=82$ to $N=90$ where the bulk part $\Delta r_{np}^\mathrm{bulk}$
remains almost constant and the surface part
$\Delta r_{np}^\mathrm{surf}$ increases roughly linearly.
 The second region, up to $N=96$, is
characterized by a fast increase of $\Delta r_{np}^\mathrm{surf}$ and,
simultaneously, a decrease of $\Delta r_{np}^\mathrm{bulk}$. Next, up to
$N=110$, $\Delta r_{np}^\mathrm{surf}$ remains roughly constant around its
maximal value reached at $N=102$ and $\Delta r_{np}^\mathrm{bulk}$ rises
almost linearly with $N$. Finally, till the shell is completely filled up at
the magic number $N=126$, $\Delta r_{np}^\mathrm{bulk}$ grows and $\Delta
r_{np}^\mathrm{surf}$ decreases, both of them linearly.

In order to explain the variations of the bulk and the surface contributions in
the NST diagram, selected sp properties of the considered Sn isotopes are
plotted in Fig. \ref{sp_sn}. First, the sp energies $\varepsilon_{nlj}$ of the
levels belonging to the major shell are shown  in panel (a) and their
occupancy $v^2_{nlj}$  in panel (b). The energy and the occupancy of these
levels show a distinct behavior  in each interval outlined above. In the first
region, from $N=82$ up to $N=90$, the lowest $2f_{7/2}$ level is
progressively populated
with neutrons. Its occupancy grows from 0 to almost 1 while  the other orbitals
remain unoccupied due to the large energy gap, fairly over 1 MeV, that
separates the $2f_{7/2}$ level from the higher levels. The $3p_{3/2}$
level crosses the
Fermi level between $N=90$ and $N=96$. Its  occupancy  increases faster in this
region than the occupancy of the $1h_{9/2}$, $2f_{5/2}$, and  $3p_{1/2}$
orbitals which are slightly higher in energy. Next, from $N=96$ to $N=110$,
the four aforementioned levels are filled  up when more neutrons are added
to the Sn nuclei.  At $N=110$ the occupancy of these orbitals ranges between 0.7
and 0.9 and in the heaviest isotopes  they are almost fully occupied.
The $1i_{13/2}$ level,  the highest in energy, behaves otherwise. Its occupancy
grows slowly up to 0.2 at $N=110$ and then it rises much faster in the heavier
isotopes, without a plateau at the shell closure at $N=126$.

We can correlate the behavior of $\Delta r_{np}^\mathrm{bulk}$ and $\Delta
r_{np}^\mathrm{surf}$  with  the principal quantum  number $n$ and the orbital
angular momentum $l$ of the orbitals occupied by the valence neutrons in the
intervals shown in Fig. \ref{R_sn}b.
The type of orbitals laying in the vicinity of the Fermi
level plays a crucial role in the determination of the neutron radii. 
When high-$n$--low-$l$ levels within the shell are populated
one observes an increase of $\Delta
r_{np}^\mathrm{surf}$, whereas low-$n$--high-$l$ levels are more
correlated with the growth of $\Delta r_{np}^\mathrm{bulk}$. 

In the considered shell of the Sn isotopes, the two $3p$ levels (displayed with
red circles in Fig. \ref{sp_sn}) have an especially large impact on the results.
At first look, it seems unexpected because only 6 neutrons out of 44 from the
whole shell can occupy these two levels. To understand this effect,
let us look at panel (c) of Fig. \ref{sp_sn} where the  mean
square radius of each sp orbital, defined as:
\begin{equation}  
\label{rnlj} 
 \lf<r^2\ri>_{nlj}= \int d{\bf r}\;  r^2   \varphi^2_{nlj}({\bf r})  \;,
\end{equation} 
is plotted for the valence shell neutrons.  In Eq. (\ref{rnlj}),
$\varphi_{nlj}({\bf r})$ is the normalized wave function.  Note that
$\lf<r^2\ri>_{nlj}$ does not depend on the occupancy $v^2_{nlj}$ or on the
multiplicity $2j+1$ of the orbital. The mean square radii of the
majority of the levels
are in the range of $30- 50$ fm$^2$ for all Sn isotopes, which is
less than twice the total mean square radii of these isotopes.
 However, the mean square radii of the $3p$
levels are much larger. They reach values of $65- 90$ fm$^2$ when
they start to
be populated at $N=90$ and of $50- 55$ fm$^2$ in the heaviest
isotopes. Such a
huge value magnifies the contribution of a few neutrons to the neutron radius
when these levels are occupied. To visualize this effect, in panel (d)
of Fig. \ref{sp_sn} we have plotted the quantity $\tau_{nlj}$, defined as:
\begin{equation}  
\label{tau1} 
\tau_{nlj}
= v^2_{nlj} \lf<r^2\ri>_{nlj}
\;, 
\end{equation} 
which describes the contribution of a single neutron to the total value of the
neutron mean square radius $\lf<r^2\ri>_n$. Thus we may write the
neutron mean square radius as
\begin{equation}  
\label{tau2} 
\lf<r^2\ri>_n
= \frac1N    \sum_{nlj} (2j+1)\tau_{nlj}\;,
\end{equation} 
where $N$ is the neutron number of the isotope.

In Fig. \ref{sp_sn}(d) we see that $\tau_{nlj}$  roughly follows the pattern of
$v^2_{nlj}$. Nevertheless, the $\tau_{nlj}$ values of the $3p$  orbitals show a
different behavior. They are magnified  in comparison to the other levels. For
example, for $N=98$ the $3p_{3/2}$  orbital has the same impact on the neutron
radius as the $2f_{7/2}$ orbital, despite their occupancies being
respectively 0.6 and 0.9.  Neutrons from the fully occupied $3p$ levels
have larger $\tau_{nlj}$ values than the other orbitals of the valence shell.
Due to the large $\lf<r^2\ri>_{nlj}$ value of the $3p$ levels, neutrons from
these orbitals  have a very strong impact on the neutron radius.

From the knowledge of the impact of particular orbitals on the structure of the
neutron surface, we can now explain the curvature of the diagrams in
Fig. \ref{R_sn}. In the first half of the shell, neutrons occupy mainly
transitional (such as $2f$) or high-$n$--low-$l$ (such as $3p$) orbitals.
They give an additional increase of the surface diffuseness of the neutron
distribution. This can be seen in Fig.~\ref{3dens}, where we compare the neutron
density profile and  its derivative of the magic nucleus $^{132}$Sn with the
mid-shell isotope $^{154}$Sn. It is clear that in $^{154}$Sn the surface
diffuseness is larger and that the slope of the surface fall-off is
smaller.
In the second half of the shell, the  influence of the low-$n$--high-$l$ level 
$1i_{13/2}$ gives the opposite effect of decreasing the surface diffuseness. 
In the particular example of the isotopes $^{132}$Sn, $^{154}$Sn, and
$^{176}$Sn we have found that $a_n$ takes values 0.526 fm, 0.815 fm and
0.720 fm, respectively.
Thus, the neutron density profile of $^{176}$Sn, also plotted in Fig.
\ref{3dens}, has a larger slope of the density distribution at the surface
than in $^{154}$Sn.

At this stage, let us summarize the above observations for the sake of
clarity. The comparison
of Figs.~\ref{R_sn}(b) and \ref{sp_sn}(d) unravels a correlation between the changes
in the isotopic shift of the NST and the sp spectrum of the orbitals in the
considered shell. When high-$n$--low-$l$ orbitals are populated (e.g. $3p_{3/2}$
between $N=90$ and 96), a rapid increase of the surface contribution $\Delta
r_{np}^{\rm surf}$ to the NST of a nucleus can be noticed. It is  manifested by
an additional increase of the NST. Conversely, when levels with
low-$n$--high-$l$ quantum numbers are being occupied (e.g. $1i_{13/2}$  above
$N=110$), the bulk
contribution $\Delta r_{np}^{\rm bulk}$ increases while $\Delta r_{np}^{\rm
surf}$ decreases. Population of orbitals with intermediate principal quantum
number and angular momentum  (such as the $2f_{7/2}$ orbital below $N=90$) supports
both the contributions to the NST.  When several types of levels are filled
simultaneously ($96\le N\le110$) we see a combination of both effects.

\begin{figure*}
\includegraphics[height=2.0\columnwidth, angle=270]{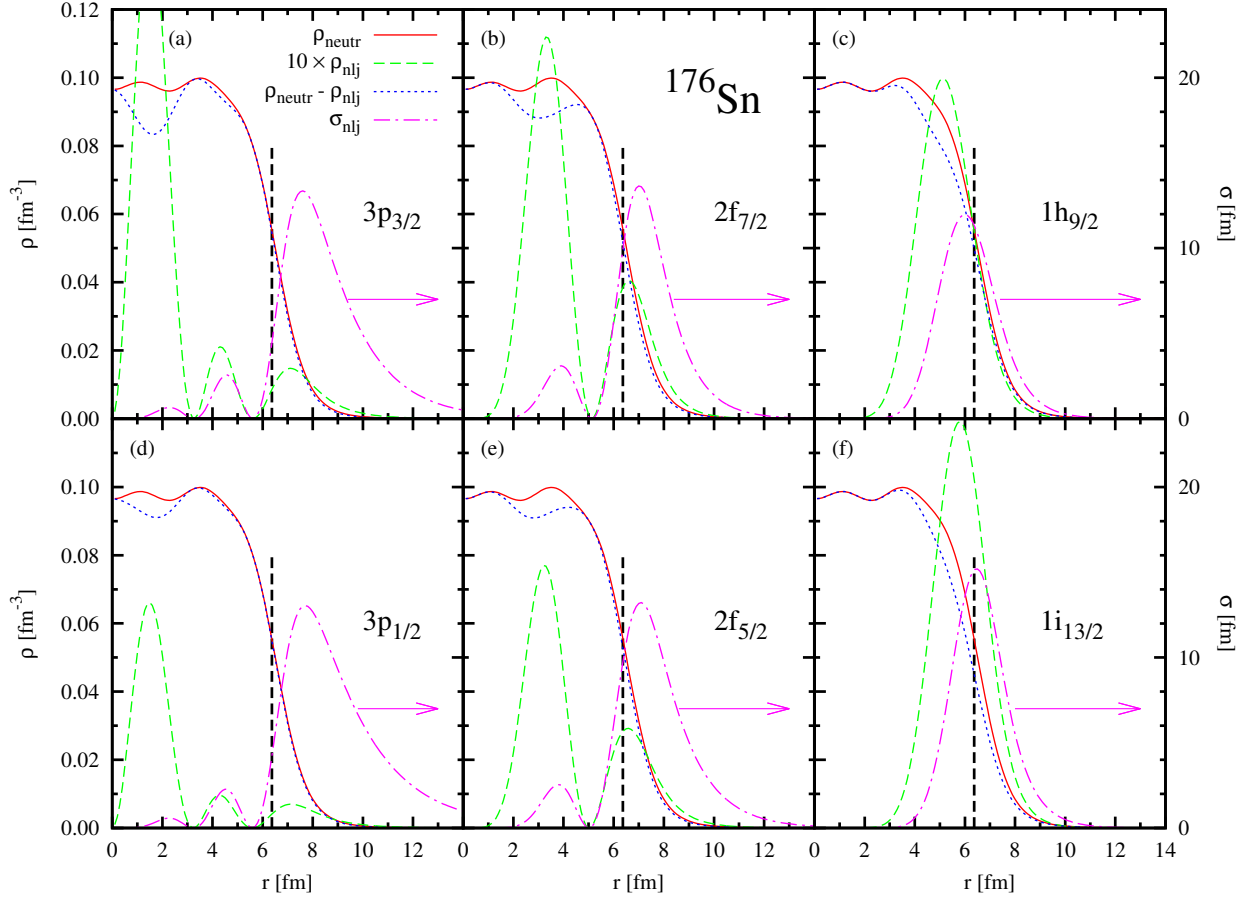}
\caption{\label{spd_sn} (Color online) Neutron density profile (red solid line) of $^{176}$Sn
compared to the single-particle density $\rho_{nlj}(r)$ of an individual orbital
defined in Eq. (\ref{rhonlj}) (multiplied by 10, green dashed line) and their
difference (blue dotted line) for the orbitals (a) $3p_{3/2}$, (b) $2f_{7/2}$,
(c) $1h_{9/2}$, (d)  $3p_{1/2}$, (e) $2f_{5/2}$, and (f) $1i_{13/2}$ as a
function of the distance from the center of the nucleus. The function
$\sigma_{nlj}(r)$ defined in Eq. (\ref{sigma1}) (magenta dot-dashed line) is
presented on the right vertical scale. The vertical dashed line indicates the
position of the half-density radius $C_n=6.367$ fm.}
\end{figure*}

\begin{figure}
\includegraphics[height=0.9\columnwidth, angle=270]{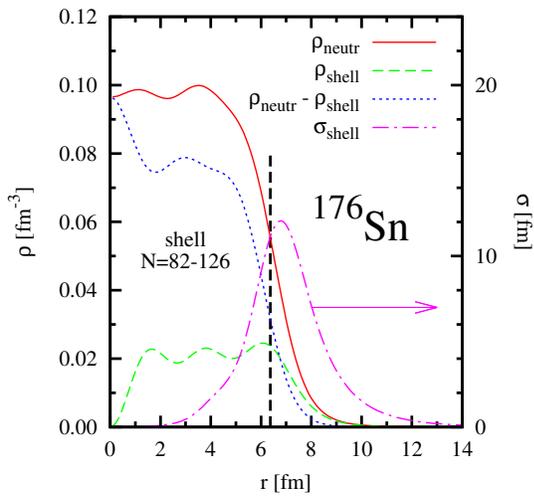}
\caption{\label{shelld_sn} (Color online) The same as in Fig. \ref{spd_sn}, but for the  whole
valence major shell from $N=82$ to $N=126$. }
\end{figure}

\begin{figure*}
\includegraphics[height=2.0\columnwidth, angle=270]{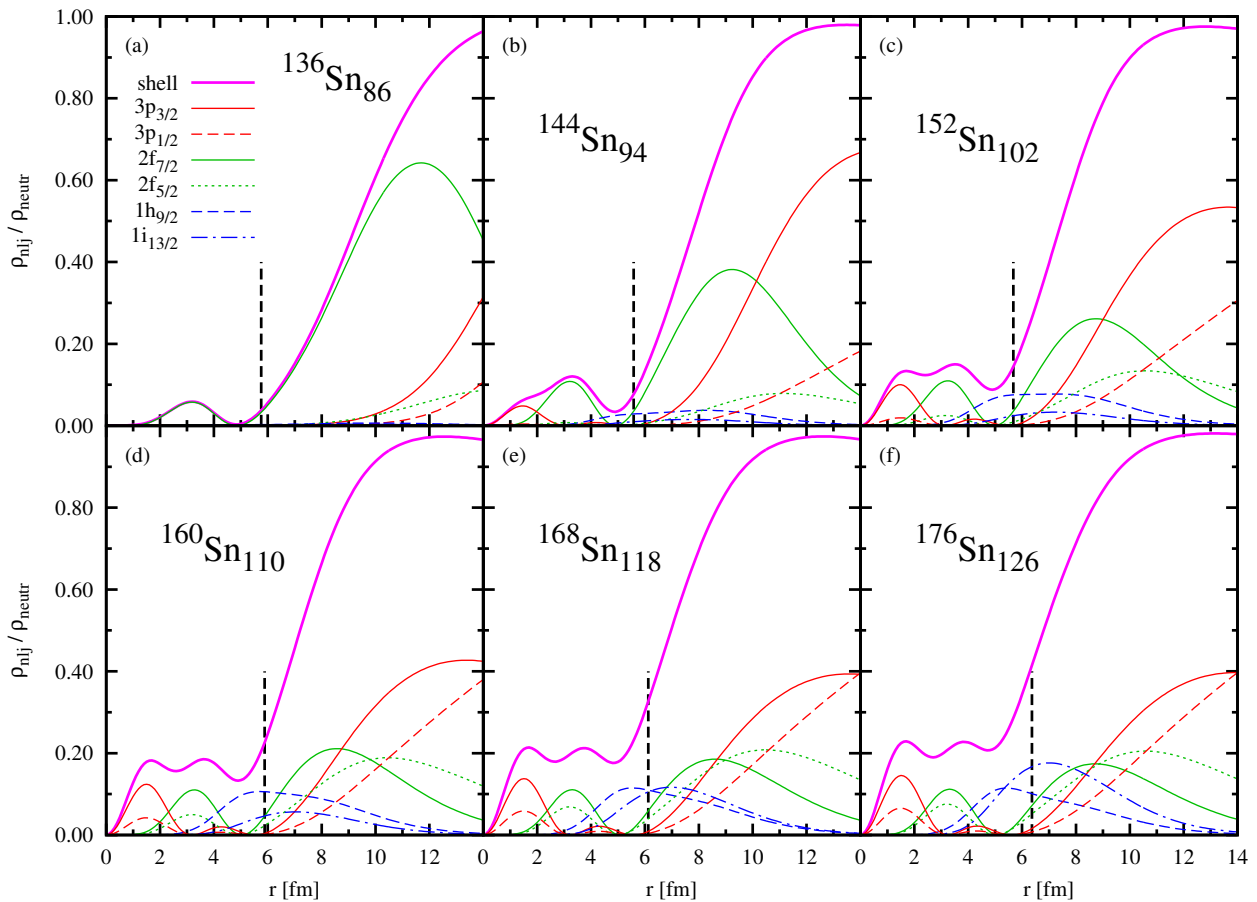}
\caption{\label{spratio_sn} (Color online) Ratio of the whole shell density (magenta thick
solid line) and of the single-particle level density to the total neutron
density as a function of the distance from the center of the nucleus in Sn
isotopes: (a) $^{136}$Sn, (b) $^{144}$Sn, (c) $^{152}$Sn, (d) $^{160}$Sn, (e)
$^{168}$Sn, and (f) $^{176}$Sn. The ratios for the $3p_{3/2}$ and $3p_{1/2}$
orbitals are marked out by red lines, the ratios for the $2f_{7/2}$ and
$2f_{5/2}$ orbitals by green lines, and the ratios for the $1h_{9/2}$ and
$1i_{13/2}$ orbitals by blue lines. The vertical dashed line indicates the
position of the half-density radius $C_n$ in each of the considered nuclei.
}
\end{figure*}

\begin{figure}
\includegraphics[height=1.0\columnwidth, angle=270]{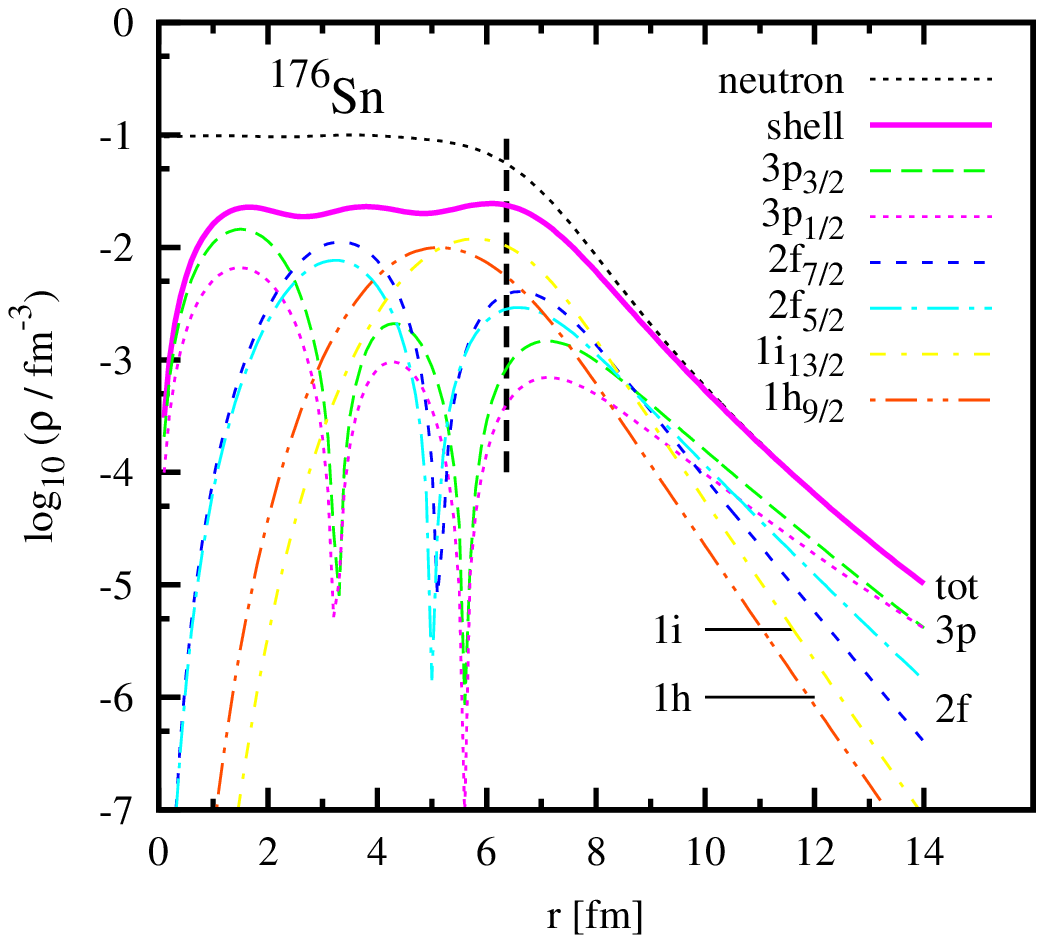}
\caption{\label{rho_log_sn} (Color online) Neutron density of $^{176}$Sn as a function of
the distance from the center of the nucleus (black dotted line) in logarithmic
scale compared to the contribution from the whole $N=82 - 126$ shell (magenta
solid line) and to the individual contributions from the  single-particle
orbitals. The densities of the orbitals are marked  out by red lines for the
$3p_{3/2}$ and $3p_{1/2}$ levels, by green  lines for the $2f_{7/2}$ and
$2f_{5/2}$ levels, and  by blue lines for the  $1h_{9/2}$ and $1i_{13/2}$
levels. The vertical dashed line indicates the  position of the half-density
radius $C_n=6.367$ fm of the nucleus.}
\end{figure}

\subsection{Single-particle neutron density distributions}

In order to find the source of the correlation described in the previous
subsection, we have to investigate  the spatial distribution of the neutrons
from each particular orbital. In Fig.~\ref{spd_sn} we analyze the neutron
distribution of $^{176}$Sn, i.e., the heaviest isotope in the considered shell
with all occupied orbitals. We compare the total neutron density
distribution, plotted  with solid lines, with the density distributions of
neutrons from each level of the valence shell (dashed  lines) defined as:
\begin{equation}
\label{rhonlj}
\rho_{nlj}(r) = (2j+1) v^2_{nlj} \varphi_{nlj}^2(r) \;.
\end{equation}
Indeed, the sp mean square radii 
defined in Eq.~(\ref{rnlj}) 
can be expressed  for
fully occupied orbitals ($v^2_{nlj}=1$)
through $\rho_{nlj}(r)$ as:
\begin{equation}
\label{rnlj2}
 \lf<r^2\ri>_{nlj}=
 \frac{4\pi}{2j+1}\int_0^\infty dr\; r^4  \rho_{nlj}(r) \;.
\end{equation}

The difference between the total density $\rho(r)$ and the density of
each orbital
$\rho_{nlj}(r)$ is also plotted in Fig.~\ref{spd_sn} with the dotted line. In
this way we find how neutrons from each level of the last shell modify
the total
neutron density profile and how they affect the neutron surface. The behavior of
the sp density distribution of each orbital is consistent with the basic quantum
mechanical properties of the nuclear orbitals.

The main contribution of the $3p$ levels to the total density (left panels of
Fig.~\ref{spd_sn}) is due to the innermost bump of  their sp densities which is
located in the interior of the nucleus at distances below 3 fm from
the
center. The two outer bumps of the density of the $3p$ orbitals
contribute much less
to the total neutron density. In the central panels of Fig.~\ref{spd_sn} we can
see that the inner bump of the $2f$ orbitals is  peaked at $r\approx 3$ fm. It
also mainly contributes to the bulk region of the total density. The
outer bump of the $2f$ orbitals,
located at the surface, is much smaller. In contrast, neutrons from
$n=1$
orbitals (right panels of Fig.~\ref{spd_sn})  practically do not contribute to
the total density at distances below 2 fm from the center. The single peak of
their density distribution is localized in the surface region with the maximum
at $r=5- 6$ fm. It can be seen that the orbitals $1h_{9/2}$ and $1i_{13/2}$
modify the density distribution by shifting the whole surface outside and
enlarging the half-density radius.

The analysis of just the density profile may not be enough when mean square
(ms) or rms radii are treated as a measure of the nuclear size. In the integral
defining the mean square radii of spherical nuclei, the nuclear density is
weighted by the fourth power of the distance from the center:
\begin{equation}
\label{r2_a}
\lf<r^2\ri>=\frac{4\pi}N \int_0^\infty dr\; r^4   \rho(r)\;.
\end{equation}
The sub-integral function in this expression is peaked at the nuclear
surface.
Therefore $\lf<r^2\ri>$ is much more sensitive to the shape of density profile
at the surface than to the nuclear bulk density.  To check the contribution of
the neutrons in each sp level to the mean square radii we have to examine the sp
densities multiplied by $r^4$. Therefore, in Fig. \ref{spd_sn} (vertical scale
on the right)  we have also plotted the function $\sigma_{nlj}(r)$ (dash-dotted
line) defined as: 
\begin{equation}
\label{sigma1}
\sigma_{nlj}(r)= 4\pi r^4  v^2_{nlj} \varphi_{nlj}^2(r)
\;.
\end{equation}
This function is the contribution of a single neutron in the $nlj$ orbital
to the mean square neutron radius that for a spherically symmetric nucleus
reads:
\begin{equation}
\label{sigma2}
\lf<r^2\ri>_n
=\frac1N  \sum_{nlj} (2j+1) \int_0^\infty dr\;  \sigma_{nlj}(r)\;.
\end{equation}
Notice that the function $\sigma_{nlj}(r)$ is related to the $\tau_{nlj}$
number, introduced before in Eq.\ (\ref{tau1}), by
\begin{equation}
\label{sigma3}
\tau_{nlj}   = \int_0^\infty dr\; \sigma_{nlj}(r)
\;.
\end{equation}

As expected, the function $\sigma_{nlj}(r)$ for all neutron orbitals is
concentrated at the surface and vanishes in the nuclear interior. The inner
bumps of $\sigma_{nlj}(r)$ in the $3p$ and $2f$ orbitals, unlike 
$\rho_{nlj}(r)$, are strongly damped in comparison with the other bumps. Some
important differences can be observed in the spatial distribution of
$\sigma_{nlj}(r)$ between the various types of orbitals. In the $3p$ levels, the
outermost bump of $\sigma_{nlj}(r)$ is peaked in the tail of the neutron density
distribution, around 1.5 fm  far from the half-density radius $C_n$ (indicated
in Fig.~\ref{spd_sn} by the black dashed vertical line). Almost the whole
contribution of $\sigma_{nlj}(r)$ to $\lf<r^2\ri>_{nlj}$ in the $3p$ levels
comes from the region outside $C_n$. The values of $\sigma_{nlj}(r)$  are
significant even at large distances beyond the nuclear surface where the neutron
density is negligible. Thus, the two $3p$ levels enlarge the neutron rms radius
mainly by modifying the surface diffuseness.

In the transitional $2f$ levels the maximum of $\sigma_{nlj}(r)$ is located
less than 1 fm outside $C_n$. Both the surface region and the tail of the
density contribute to the function $\sigma_{nlj}(r)$. The influence of the
$2f$ orbitals on the neutron skin is more ambiguous than in the case of the $3p$
orbitals because they simultaneously contribute to the surface part and to the
bulk part of the NST.

The function $\sigma_{nlj}(r)$ of the low-$n$--high-$l$ neutrons ($1h_{9/2}$ and
$1i_{13/2}$) has a single bump with a maximum in the vicinity of the
half-density radius $C_n$. It is distributed rather symmetrically around $C_n$
and does not extend to larges distances. Neutrons from these orbitals almost do
not modify the surface diffuseness $a_n$, but increase the half-density radius
$C_n$. As a consequence, the surface contribution $\Delta r_{np}^{\rm surf}$ to
the neutron skin diminishes and the bulk part $\Delta r_{np}^{\rm bulk}$
increases substantially.

In Fig.~\ref{shelld_sn} we have plotted for $^{176}$Sn the same functions as in
Fig.~\ref{spd_sn} but for all neutrons  from the whole valence shell.  The net
total contribution of the neutrons from the last shell, denoted  by $\rho_{\rm
shell}(r)$, is displayed by the dashed line in Fig.~\ref{shelld_sn}. It is
distributed rather evenly throughout the nuclear interior. Due to the lack
of $s$ orbitals in the considered shell, it vanishes in the center of the
nucleus.
The total density distribution in the surface region is determined mostly by the
neutrons  from the last shell. The contribution of $\sigma_{\rm shell}(r)$ to
the neutron mean square  radius averages the contributions of all the orbitals
of the valence shell. It is peaked around the half-density radius and it has
quite a large width, extending to the area outside the nuclear surface.
Therefore, the shell as a whole contributes to both the bulk and surface parts
of the neutron skin.

\begin{figure*}
\includegraphics[height=1.6\columnwidth, angle=270]{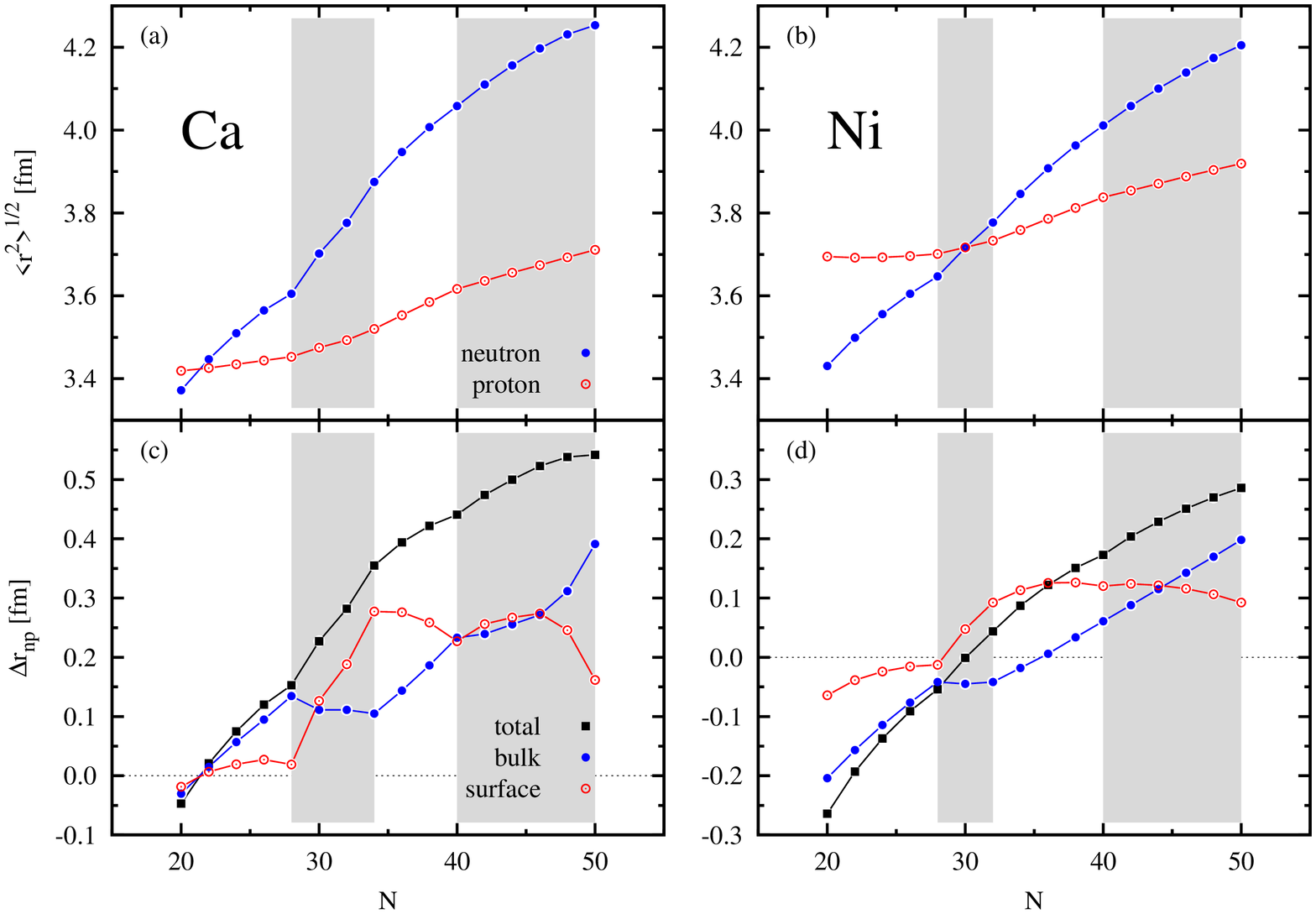}
\caption{\label{R_ca} (Color online) The same as in Fig. \ref{R_sn} but for Ca  and Ni isotopes with
$20\le N\le 50$.}
\end{figure*}
\begin{figure*}
\includegraphics[width=1.6\columnwidth]{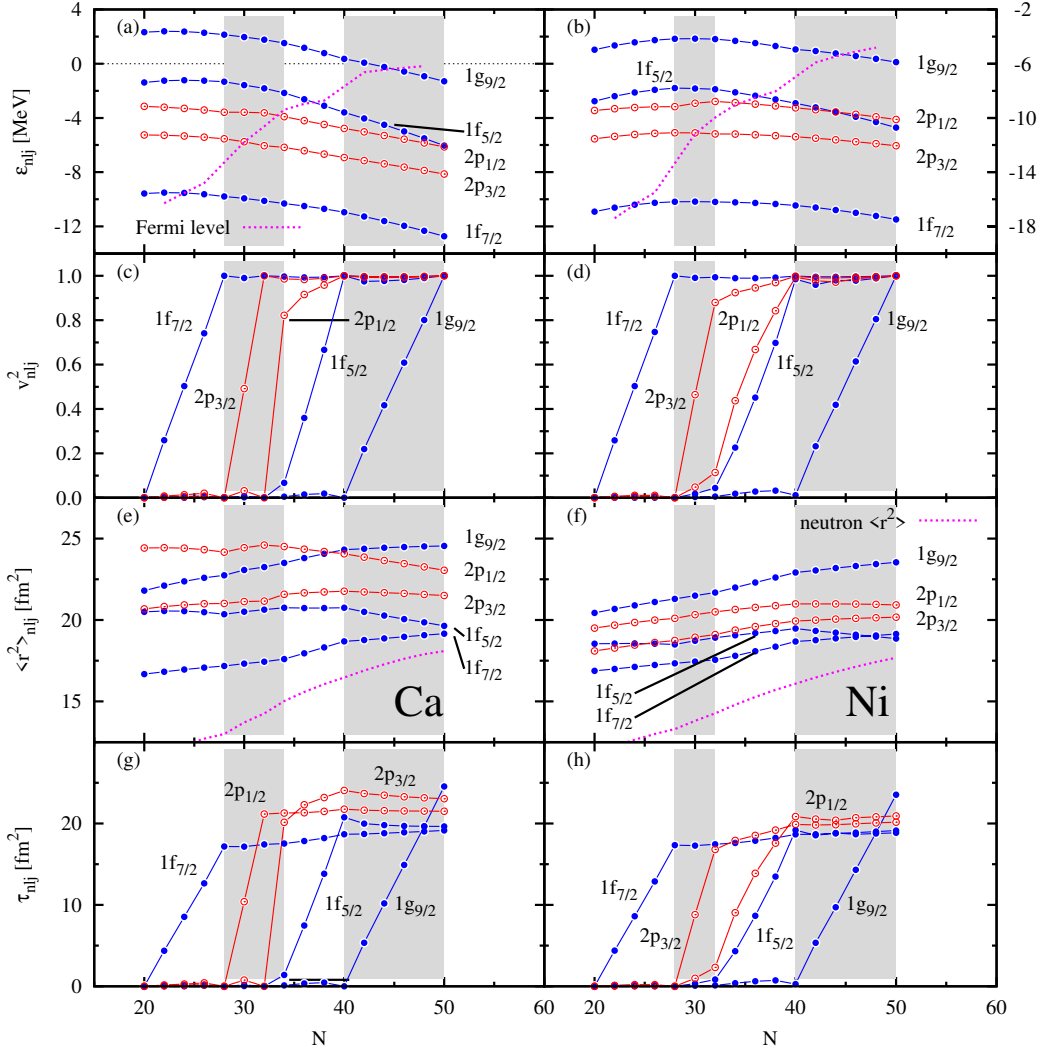}
\caption{\label{sp_ca} (Color online) The same as in Fig. \ref{sp_sn} but for Ca and Ni isotopes with
$20\le N\le 50$. Note that the vertical scale of panel (b) is on the right axis.
}
\end{figure*}
\begin{figure*}
\includegraphics[height=2.0\columnwidth,angle=270]{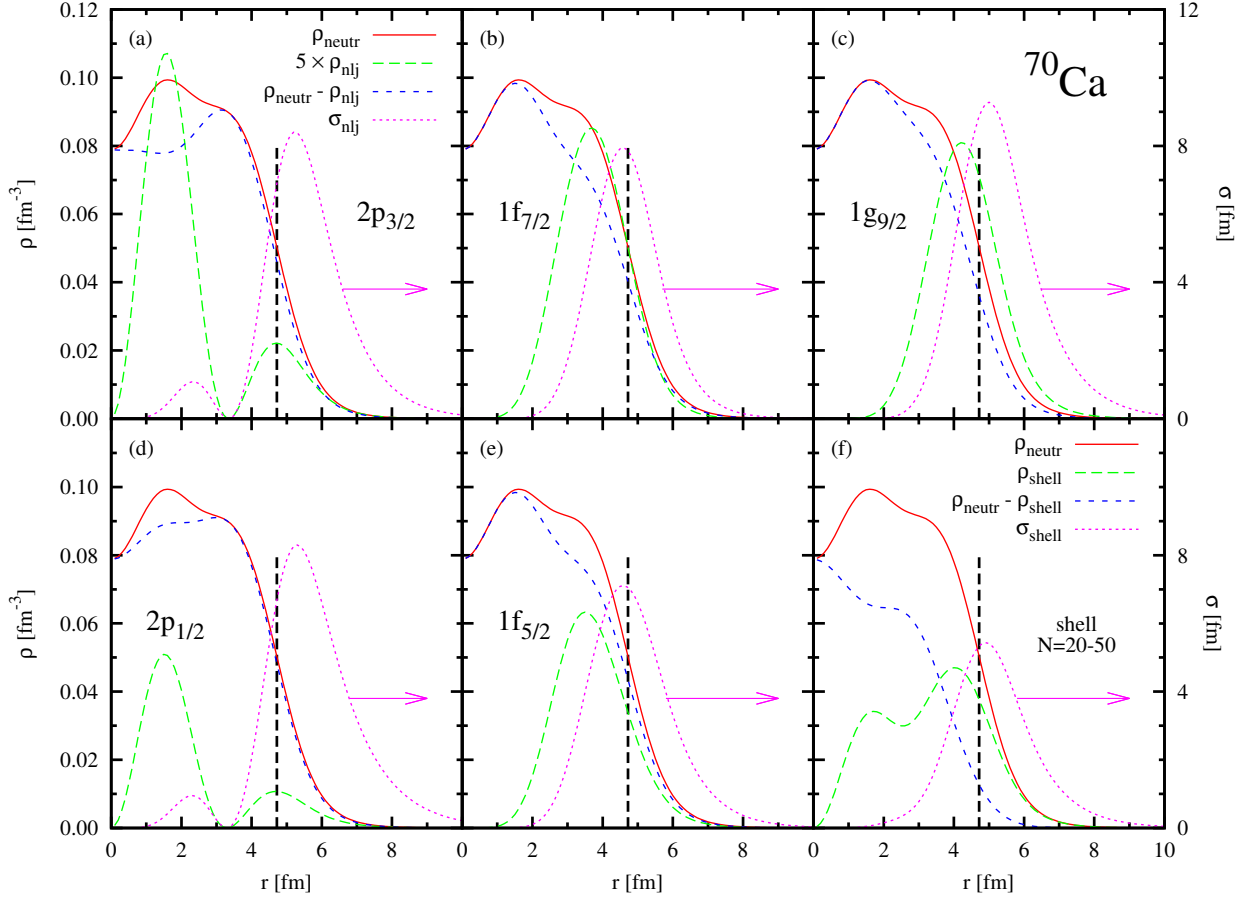}
\caption{\label{spd_ca} (Color online) The same as in Figs. \ref{spd_sn} and \ref{shelld_sn}
but for $^{70}$Ca in panels (a)-(f) and for the $20\le N\le 50$ shell of Ca in
panel (g).}
\end{figure*}

\begin{figure*}[t]
\includegraphics[height=1.6\columnwidth, angle=270]{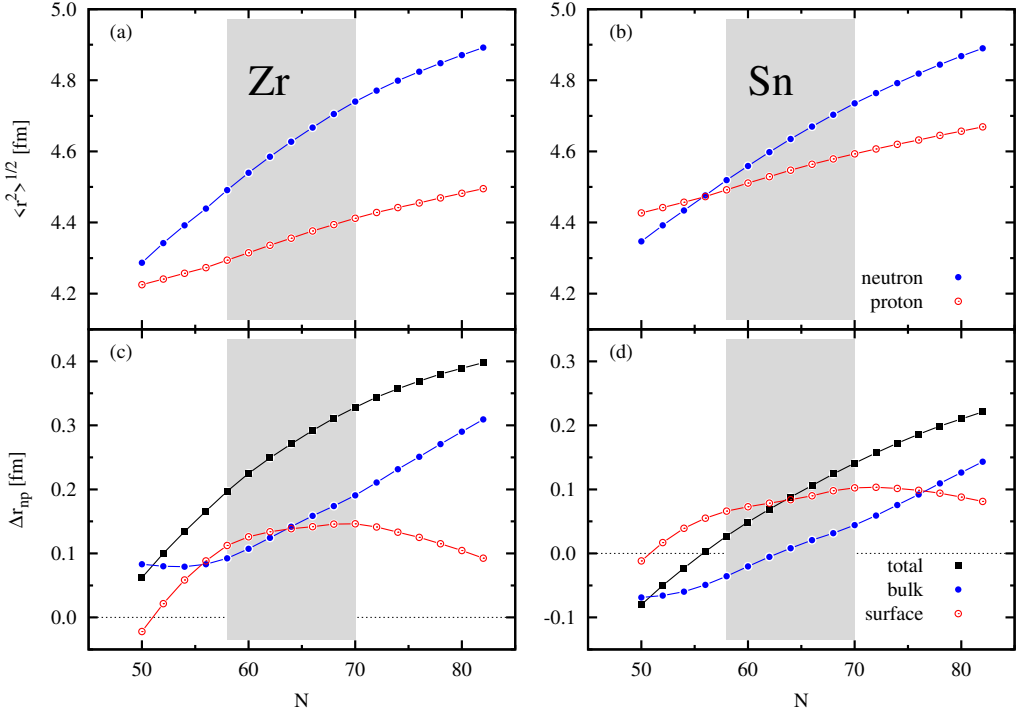}
\caption{\label{R_zr} (Color online) The same as in Fig. \ref{R_sn} but for Zr and Sn isotopes with
$50\le N\le 82$.
}
\end{figure*}
\begin{figure*}
\includegraphics[width=1.6\columnwidth]{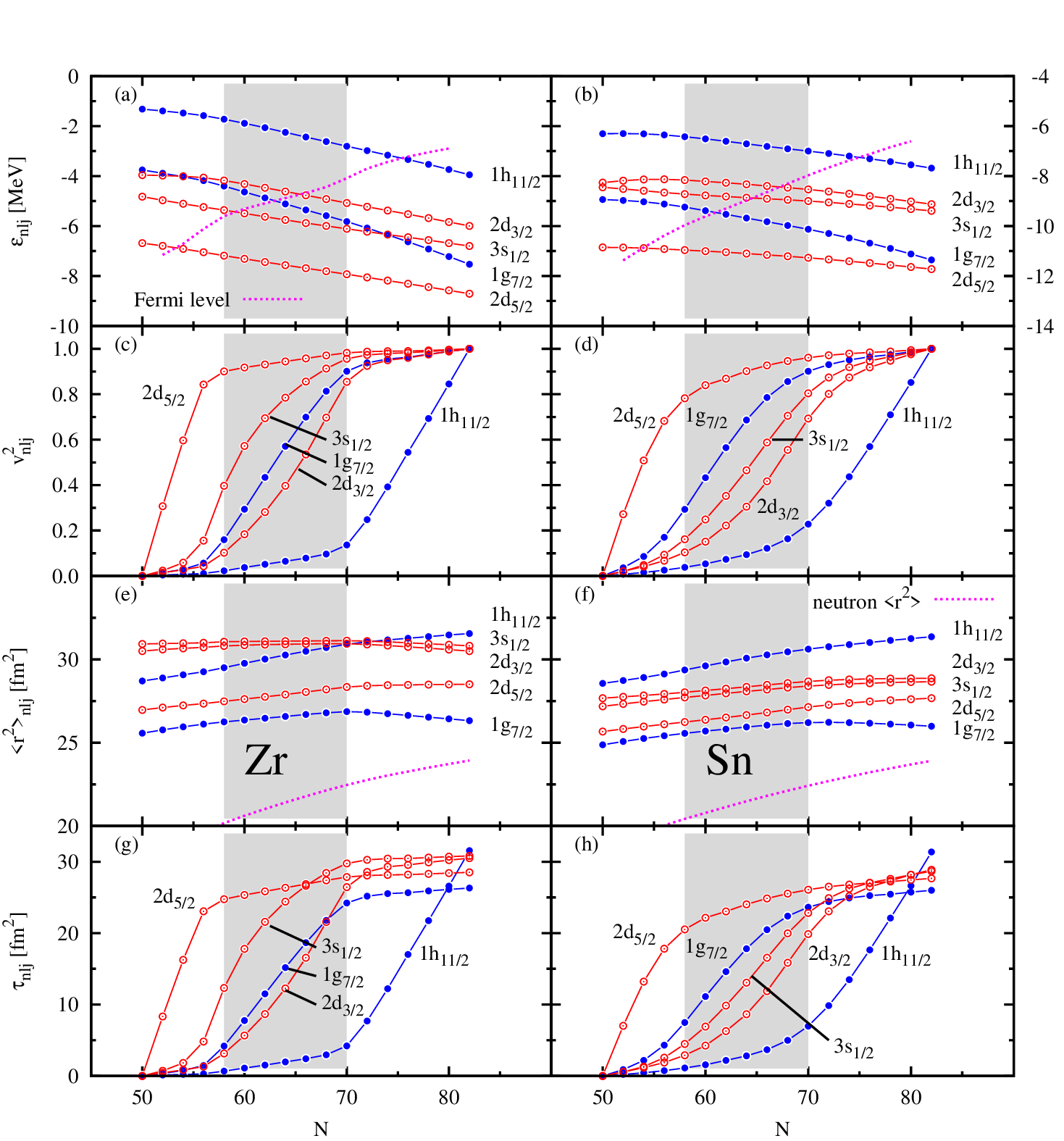}
\caption{\label{sp_zr} (Color online) The same as in Fig. \ref{sp_sn} but for Zr
and Sn isotopes with
$50\le N\le 82$. Note that the vertical scale of panel (b) is on the right axis.}
\end{figure*}

\begin{figure}
\includegraphics[height=0.9\columnwidth, angle=270]{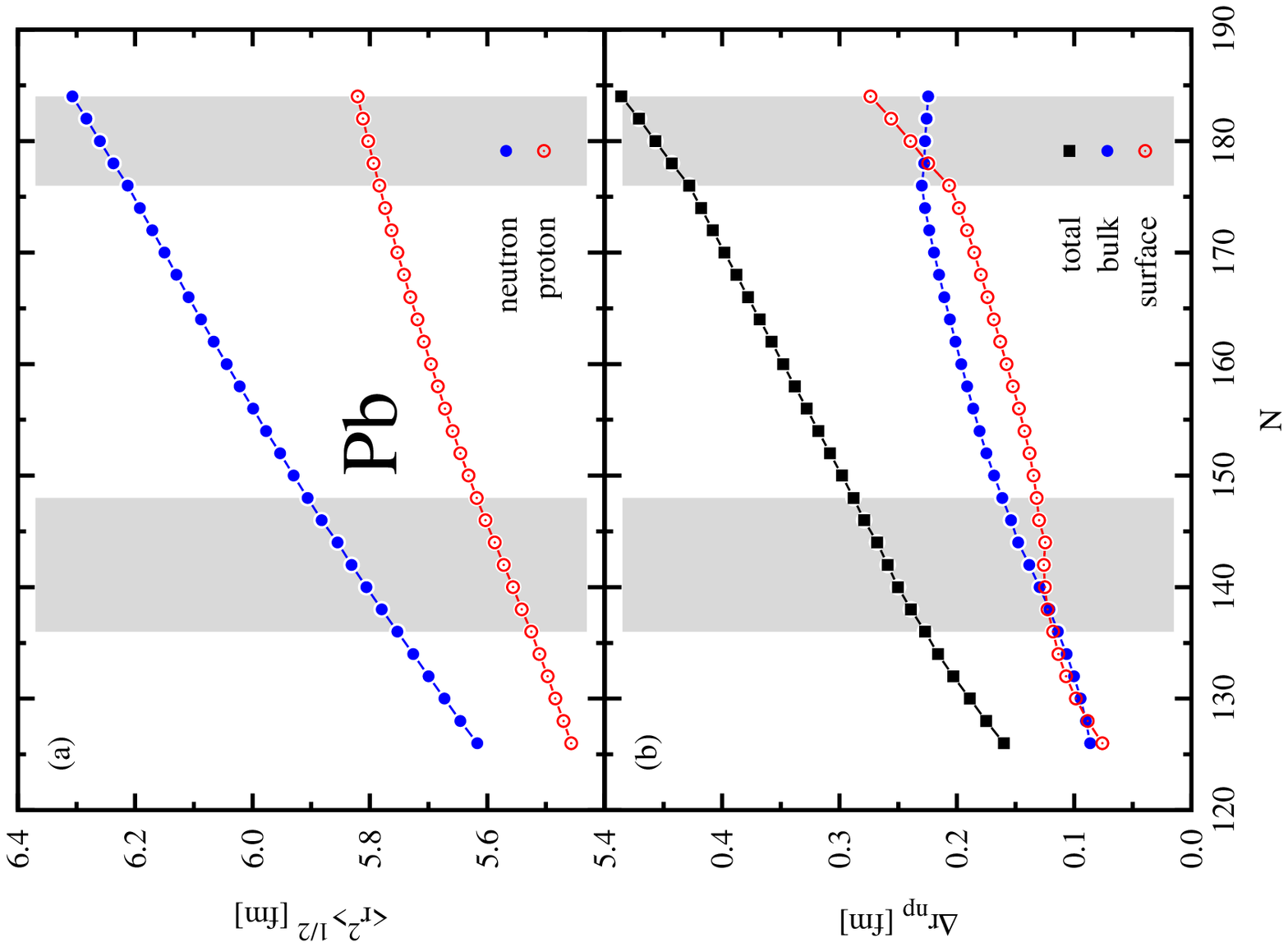}
\caption{\label{R_pb} (Color online) The same as in Fig. \ref{R_sn} but for Pb isotopes with
$126\le N\le 184$.
}
\end{figure}
\begin{figure}
\includegraphics[width=0.9\columnwidth]{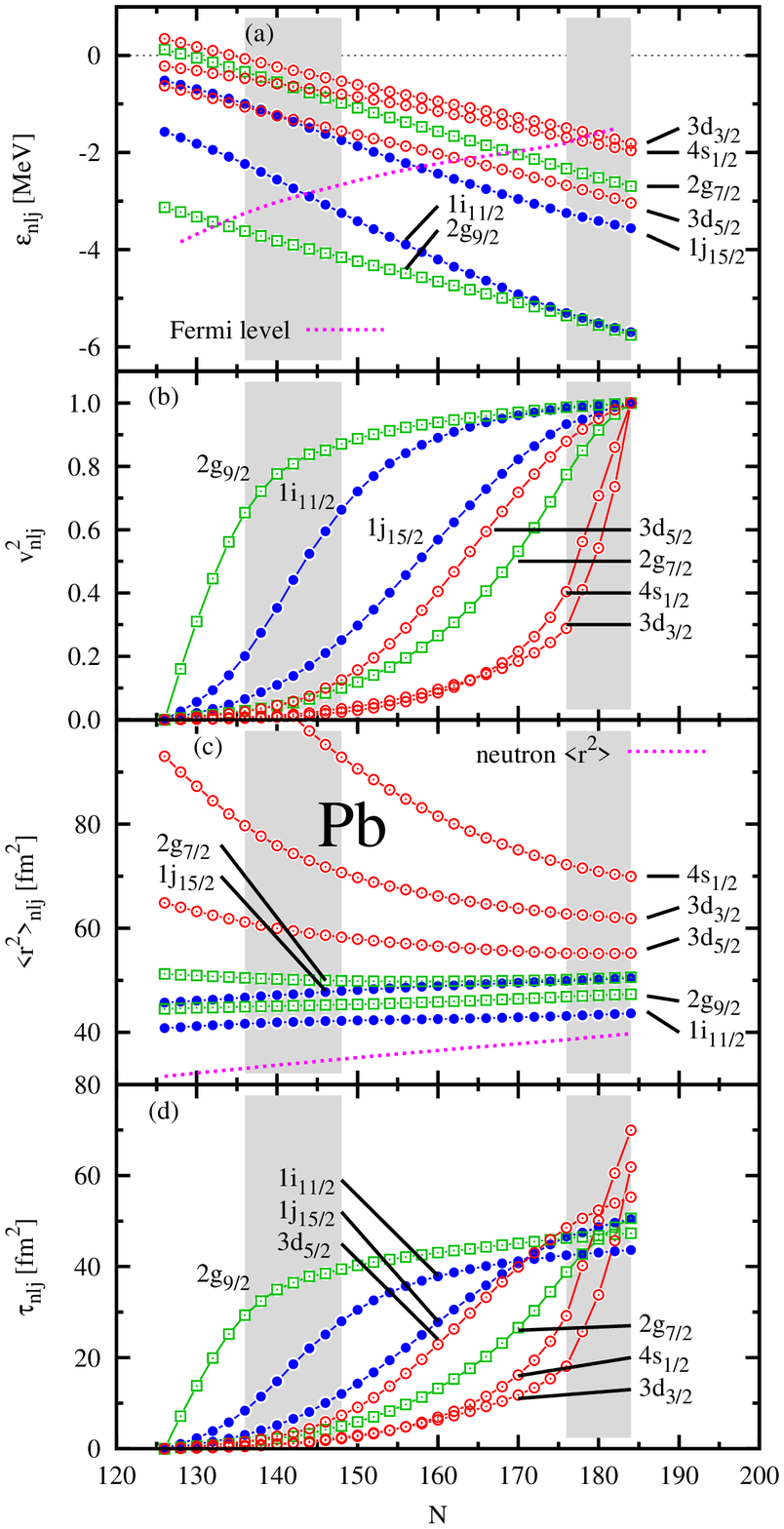}
\caption{\label{sp_pb} (Color online) The same as in Fig. \ref{sp_sn} but for Pb isotopes with
$126\le N\le 184$.}
\end{figure}

\subsection{Structure of the tail of the neutron density distribution}

In the previous subsection, we have found that the tail of the density beyond
the mean position of the nuclear surface gives a non-negligible contribution to
the neutron mean square radii and is significant for the neutron skin formation.
Now we will discuss the structure of the tail of the neutron densities along the
valence shell of the Sn isotopic chain.
In Fig.\ \ref{spratio_sn} we have plotted with the thick solid line the ratio of
the density of the nucleons from the whole $N= 82 - 126$ shell to the total
neutron density
$\rho_{\rm shell}(r)/\rho_{\rm neutr}(r)$ for the few selected Sn isotopes along the
considered shell. In all these nuclei, $\rho_{\rm shell}(r)/\rho_{\rm neutr}(r)$ grows
from $r= 5$ fm and reaches
90\% at $r= 10 - 12$ fm. Hence, the valence nucleons
play a dominant role in the structure of the tail of the density distribution. 

In Fig.\ \ref{spratio_sn} the ratio $\rho_{nlj}(r)/\rho_{\rm neutr}(r)$ of the orbital
density, defined in Eq.\ (\ref{rhonlj}), to the total neutron density is
also plotted \cite{bar97}. The huge influence of the single neutrons on the tail
of the nuclear density  distribution can be noticed. In panel (a) of Fig.\
\ref{spratio_sn}, the case of $^{136}$Sn is shown, where only 4 neutrons mainly
from the $2f_{7/2}$ orbital are present in the last shell. The ratio to the
total density reaches around 60\% at $r=10$ fm. The contribution of the
$2f_{7/2}$ level in heavier
isotopes diminishes in favor of the other orbitals, but even in the heaviest
isotopes its maximum remains at the level of a 20\% contribution. In all of the
isotopes, neutrons from only a few orbitals determine the total density of the
nuclear tail.
Even as close to the bulk as at half-density radius, selected orbitals share
$10\% - 20 \%$ of the total density, e.g. $1i_{13/2}$ in $^{176}$Sn.
Despite they accomodate only 6 neutrons, the $3p_{3/2}$ and $3p_{1/2}$
levels seem to be dominant in
the tail at large distances, where these two levels together account for
even 80\% of the neutron density.

In the analysis of the sp contribution to the tail of the neutron
density we cannot forget about the rapid, exponential decrease of the
neutron density at large distances outside the nuclear surface. It is depicted
in Fig.~\ref{rho_log_sn} where the neutron density of the nucleus $^{176}$Sn is
plotted in logarithmic scale. It can be noticed that beyond
$r=10$ fm the neutron density of $^{176}$Sn decreases below
$10^{-3}$ fm$^{-3}$, which is 1\% of the bulk density. At these distances only
neutrons from the valence shell $N=82- 126$ contribute to the
total neutron density, which confirms the conclusions deduced from Fig.\
\ref{spratio_sn}f. When we look into the individual 
densities of the sp orbitals,  also displayed in Fig.~\ref{rho_log_sn}, we
notice that the type of orbital determines the slope of the logarithmic decrease
of the sp neutron density. The fall-off of the $3p$ levels is slower than for
the $2f$ levels and the largest slope is found for the $1h_{9/2}$ and
$1i_{13/2}$ levels. Thus, the $3p$ levels participate much more than the
low-$n$--high-$l$ levels to the total neutron density at very large distances
from the center of the nucleus, as expected from their high mean square radii
[cf. $\lf<r^2\ri>_{nlj}$ in panel (c) of Fig.~\ref{spd_sn} and $\sigma_{nlj}(r)$
in Fig.~\ref{shelld_sn}]. At distances larger than about 4 fm from the 
half-density radius, the $3p$ levels determine the slope of the fall-off of the
neutron  density. We have to remember that in this region the $3p$ orbitals
still contribute to the neutron rms radii as the densities are weighted by $r^4$
in the calculation of this observable [cf.\ Eq.~(\ref{r2_a}) and Figs.
\ref{spd_sn} and \ref{shelld_sn}]. The shape of the tail of the nuclear density
is also very important in the analysis of the experiments performed in exotic
atoms such as antiprotonic or pionic atoms. The heavy, negative charged particle
annihilates at distances around 2 or 3 fm from the nuclear surface
\cite{klo07,fri05,fri09} and it is highly sensitive to the differences between
the neutron and proton density distributions in this region.

In this Section we have studied the correlation between the quantum numbers of
the valence neutrons and the neutron skin of Sn isotopes. Neutron orbitals with
low principal quantum number $n$ and high angular momentum $l$ are localized at
the nuclear surface. They induce a shift of the neutron half-density radius and
increase the bulk contribution to the NST. Neutron levels with high principal
quantum number $n$ and low angular momentum $l$ can be found in the
same shell. The outermost bump of their density distribution is localized in the
tail of the neutron density profile. It brings about a large contribution to the
mean square neutron radii and determines the slope of the logarithmic fall-off
of the neutron density outside the surface region. The high-$n$--low-$l$
orbitals are responsible for increasing the surface contribution to the NST.


\section{Neutron skin in the other isotopic chains}


With the experience gained in the previous Section, we analyze on the same
footing the other isotopic chains with magic proton number, namely Ca, Ni, 
Zr, light Sn isotopes
and Pb, which are representative of different mass regions.

\subsection{Ca and Ni isotopes with $20\le N \le 50$ }

We study Ca and Ni isotopes in the range covering the two major neutron shells
$N=20- 28$ and $N=28- 50$. The neutron drip line of the Ca isotopes
is reached at $N=48$ close to the upper limit of the $N=28- 50$ shell,
whereas it is shifted to $N=60$ for Ni.
In Fig.\ \ref{R_ca} (a) and (b) we can see, similarly to the case of the
Sn isotopes, that the proton rms radius of the Ca and Ni isotopic chains
grows rather smoothly, whereas the neutron rms radius increases with a
variable slope. The additional 8 protons in Ni make the proton rms radii
of this element larger than in Ca by $\sim\!0.25$ fm.
However, the neutron rms radii change very little from Ca to Ni. As a
consequence, the NST in the Ni isotopes is around 0.25 fm smaller than in Ca
for the same number of neutrons. Indeed, in the neutron-deficient isotopes
of Ni from $N=20$ till $N=28$ the NST [as well as its bulk and surface
parts, which are plotted in Fig.\ \ref{R_ca} (c) and (d)] is negative. The
total change in absolute value of the NST from $N=20$ to $N=50$ is similar
in both the Ni and Ca isotopic chains, that is, around 0.55 fm for Ni and 0.59
fm for Ca.

Looking at the bulk $\Delta r_{np}^\mathrm{bulk}$ and surface $\Delta
r_{np}^\mathrm{surf}$ contributions to the NST in Ca and Ni, one sees from
Fig.\ \ref{R_ca} that the largest part of the change of the NST from Ca to
Ni comes from the change of $\Delta r_{np}^\mathrm{bulk}$ between these
two elements, whereas $\Delta r_{np}^\mathrm{surf}$ changes less from one
element to the other element. In fact, in the case of the magic numbers $N=20$,
$N=28$, and $N=50$ it is seen that the value of $\Delta r_{np}^\mathrm{surf}$ is
almost equal in Ca and Ni. Thus, the decrease of the NST from Ca to Ni
in these nuclei with magic neutron number is produced almost entirely by
the decrease of the bulk contribution $\Delta r_{np}^\mathrm{bulk}$.

Four intervals of the neutron number are clearly distinguished in the bulk and surface 
contributions to the NST in Fig.\ \ref{R_ca} (c) and (d). In the first interval, corresponding
to the $N=20- 28$ shell, the bulk contribution to the NST grows almost
linearly and the surface contribution remains practically constant. 
The low-$n$--high-$l$ level $1f_{7/2}$ is populated within this region, as
it can be seen from panels (a)--(d) of Fig.\ \ref{sp_ca} which display the
position and occupation of the levels close to the Fermi energy for the Ca and Ni isotopes.
The second interval of the neutron number
ranges from $N=28$ to $N=34$ in Ca and to $N=32$ in Ni. In these nuclei the
surface contribution to the NST grows rapidly while the bulk part slightly
decreases. The $2p_{3/2}$  level is
being filled mostly in this interval as well as the $2p_{1/2}$ level in
Ca [cf.\ Fig.\ \ref{sp_ca} (a)--(d)].

The third interval of Fig.\ \ref{R_ca} covers the isotopes up to
$N=40$ where the bulk part of the NST increases and the surface part 
slightly decreases (Ca) or remains almost constant (Ni).
Within this region, in Ca mainly the $1f_{5/2}$ level is being occupied. The  $2p_{1/2}$ level lies closer to the $1f_{5/2}$ level in Ni than in
Ca, and consequently, these two orbitals in Ni are populated almost
simultaneously between $N=32$ and $N=40$ [cf.\ Fig.\ \ref{sp_ca} (a)--(d)].
Thus, the third interval in neutron number of the Ni isotopes
starts already at $N=32$, rather than at $N=34$ as in Ca.
In this third interval of Ni the influence of the
$2p_{1/2}$ orbital on the NST of Ni is overpowered by neutrons from the
$1f_{5/2}$ orbital; one finds a very modest rise of the
surface part $\Delta r_{np}^\mathrm{surf}$ of the NST while the bulk part
$\Delta r_{np}^\mathrm{bulk}$ grows linearly (see Fig.\ \ref{R_ca}d).

The fourth interval of Fig.\ \ref{R_ca} ranges between $N=40$ and $N=50$, where occupation of the
$1g_{9/2}$ level takes place. In Ca between $N=40$ and $N=46$ both the bulk and
surface contributions to the NST take a similar value with a slight growing
trend with $N$. For larger number of neutrons in the Ca isotopes, $\Delta r_{np}^\mathrm{bulk}$
increases faster again whereas $\Delta r_{np}^\mathrm{surf}$ decreases till
reaching $N=50$. In Ni, from the semimagic number $N=40$ till $N=50$
a clear linear increase of $\Delta r_{np}^\mathrm{bulk}$ is accompanied by a
slow decrease of $\Delta r_{np}^\mathrm{surf}$. This behavior is in agreement with the
general trend for the low-$n$--high-$l$ orbitals, as in Ni only the $1g_{9/2}$ level is populated.

The source of the changes in the behavior of the NST becomes clear in
Fig.\ \ref{sp_ca} where the sp properties of the valence orbitals are displayed.
We can discuss them relying on the analysis performed in the previous Section
for Sn isotopes. In Fig.\ \ref{sp_ca} it can be seen that the neutron orbitals
in the vicinity of the Fermi level are rather well separated in energy. Hence,
we expect quite clear differences in $\Delta r_{np}^\mathrm{bulk}$ and
$\Delta r_{np}^\mathrm{surf}$ between neutron number intervals.

The $2p$ orbitals of Ca and Ni play a similar role to the $3p$ levels in Sn. They
should contribute mainly to the surface of the neutron distribution. Indeed, a
sudden increase of $\Delta r_{np}^\mathrm{surf}$ is observed in the second
interval of Fig.\ \ref{R_ca}.
It can be noticed from Fig.\ \ref{sp_ca} that in Ca and Ni the mean square
radii of the $2p$ levels are similar to those of the neighboring levels,
unlike the case of the $3p$ orbitals in the Sn isotopes.
The total neutron density and the sp densities $\rho_{nlj}(r)$ of the individual
neutron orbitals defined in Eq.\ (\ref{rhonlj}), as well as the function
$\sigma_{nlj}(r)$ defined in Eq.\ (\ref{sigma1}), are plotted in  Fig.\
\ref{spd_ca} for the $^{70}$Ca nucleus.
Though the peak of the $\sigma_{nlj}(r)$ function of the $2p$ levels of
Ca lies closer to the half-density radius than in the case of the $3p$ levels of
Sn (cf.\ Fig. \ref{spd_sn}), the $2p$ levels still give an important
contribution to the tail of the neutron density.

The low-$n$--high-$l$ levels $1f_{7/2}$, $1f_{5/2}$, and $1g_{9/2}$ are
expected to contribute to the NST by increasing the bulk part. In
fact, in the first, third, and fourth intervals of Fig.\ \ref{R_ca} we see a
clear increase of $\Delta r_{np}^\mathrm{bulk}$. In the Ca chain
the last low-$n$--high-$l$ orbital $1g_{9/2}$ is bound only for $N>42$ with
energy close to zero, as seen in panel (a) of Fig.\ \ref{sp_ca}. The continuum
affects the behavior of the Ca $1g_{9/2}$ orbital at large distances from
the nuclear center.
This almost unbound orbital extends to a region far away from the nucleus.
Consequently, the tail of the $\sigma_{nlj}(r)$ function for the $1g_{9/2}$
level (see panel (c) of Fig.~\ref{spd_ca}) is larger than for the $1f$
levels (central panels of Fig.~\ref{spd_ca}). Hence, a relatively
large impact of the $1g_{9/2}$ level in the surface contribution to the NST
of the Ca isotopes beyond $N=40$ is observed in Fig.~\ref{R_ca}c.
On the other hand, in Ni isotopes the $1g_{9/2}$
level is shifted down by around 5 MeV in comparison to Ca
owing to the larger number of protons in Ni. Its energy is far below zero and
therefore in Ni the neutron skin splitting in bulk and surface parts
for this low-$n$--high-$l$ orbital is not disturbed by the influence of the
continuum, unlike in Ca.

\subsection{Zr and Sn isotopes with $50\le N\le 82$}

We have considered the shell with $N=50- 82$ neutrons in the Zr and Sn
elements \cite{vin12}. 
The neutron drip line of the Zr isotopes is reached at the end of this range.
In Fig. \ref{R_zr} we can see again the roughly
linear increase of the proton rms radii with larger $N$, whereas the
neutron rms radii, as well as the NST, rise with a rather curved
shape.
We can also see that the difference in proton number between the two
elements affects mainly the proton rms radii, shifting them by a
practically constant value of around 0.2 fm along the shell. The neutron
rms radii, however, remain very similar in the two elements. As a
consequence, the NST decreases from Zr to Sn by the same amount as the
proton radii increase from Zr to Sn.

If we compare the values of the bulk and surface contributions to the NST in Zr with the
values of these contributions in Sn, it turns out
that the bulk contribution $\Delta r_{np}^\mathrm{bulk}$ has decreased from Zr
to Sn by a roughly constant shift of 0.15 fm, whereas the surface
contribution $\Delta r_{np}^\mathrm{surf}$ remains almost the same or
decreases by no more than 0.05 fm from Zr to Sn. Altogether, it implies that the
change of the NST between the two elements takes place, in essence,
through the modification of the sharp radius of the proton density
distribution caused by the different proton number, whereas the surface
diffuseness remains less affected. Actually, it is interesting to observe
that in the nuclei with neutron magic number ($N=50$ and $N=82$) the
value of $\Delta r_{np}^\mathrm{surf}$ is practically identical in Zr and
Sn, which means that in these nuclei the change of $\Delta r_{np}$ between Zr
and Sn is exclusively due to the change of $\Delta r_{np}^\mathrm{bulk}$.
These observations are in consonance with what we had found before in the study
of the NST in Ca and Ni.

The bulk and the surface contributions to the NST in the Zr and Sn isotopes,
unlike in the previously discussed Ca and Ni elements, change their slopes rather
smoothly without any kinks in the graph [see Fig. \ref{R_zr} (c) and (d)]. Nevertheless, three intervals can be
distinguished.  In the first interval, up to $N=58$, the surface part
grows linearly.  In the second interval with
$N=58- 70$, the surface contribution  is roughly constant around its maximal
value. In the third interval, beyond $N=70$, $\Delta r_{np}^\mathrm{surf}$
decreases. The bulk contribution remains practically constant up to $N=58$ and
for heavier isotopes it increases linearly up to the end of the shell.

Five levels from the major shell
with $N=50- 82$ can be found in the energy spectrum of the Zr and Sn isotopes,  as
displayed in  Fig. \ref{sp_zr}. There are two low-$n$--high-$l$ levels:
$1g_{7/2}$ and $1h_{11/2}$, and three high-$n$--low-$l$ levels: $2d_{5/2}$,
$2d_{3/2}$, and $3s_{1/2}$. In the first interval till $N=58$, the $2d_{5/2}$
level is being
filled almost without interference from the other orbitals [see panels (c) and (d)
of Fig. \ref{sp_zr}]. Similarly, in the last interval beyond $N=70$ mainly the
$1h_{11/2}$ level is being populated. It explains the behavior shown in Figs.
\ref{R_zr}c and \ref{R_zr}d  by the bulk and
surface contributions to
the NST in the regions between $N=50-58$  and $N=70-82$ of the Zr and Sn isotopic
chains.
In the intermediate region between $N=58$ and $N=70$, the contribution of
the $2d_{5/2}$ level almost saturates and the  $1h_{11/2}$ level practically
does not yet contribute to the neutron radii. The remaining 
$3s_{1/2}$, $2d_{3/2}$, and $1g_{7/2}$ levels play the main role in this interval.
The influence of the $3s_{1/2}$
and $2d_{3/2}$ orbitals keeps $\Delta r_{np}^\mathrm{surf}$ from falling down,
whereas the presence of the $1g_{7/2}$ orbital makes
$\Delta r_{np}^\mathrm{bulk}$ to increase.

\subsection{Pb isotopes with $126\le N\le  184$ }

The last element considered in this article is Pb. We concentrate on
neutron-rich isotopes with $126\le N\le 184$. 
The Pb neutron drip line is reached at the end of this shell.
In Fig. \ref{R_pb} we can see an almost linear behavior of
the rms radii for both protons and neutrons. Also the NST and its bulk and
surface contributions are overall roughly linear with $N$. As it can be seen in
Fig. \ref{sp_pb}, the large level density around the Fermi energy in these heavy
nuclei causes the occupancy of all orbitals to rise along the whole shell.
Hence, the influence of the sp properties onto the nuclear surface is smoothed
out in this element. Nevertheless, the slight changes of the slope  of the bulk
and surface contributions to the NST, which can be seen in panel (b) of Fig.
\ref{R_pb}, allow one to divide the shell into four regions.
In order to explain these variations we
shall look into the orbitals from this valence shell, which
are displayed in Fig. \ref{sp_pb}. There are seven different levels in this
shell. We classify them into three groups: the low-$n$--high-$l$ levels
$1i_{11/2}$ and $1j_{15/2}$, the transitional levels $2g_{9/2}$ and $2g_{7/2}$,
and the high-$n$--low-$l$ levels $3d_{5/2}$, $3d_{3/2}$, and $4s_{1/2}$.
Although the occupancy of all levels changes throughout
the shell, in each interval the occupancy of some orbitals increases much faster
than the occupancy of the other orbitals.

In the first interval, up to $N=136$, the bulk and the
surface parts of the NST increase with a similar slope. The occupancy of the
transitional $2g_{9/2}$ level grows very fast here.
In the second region, between $N=136$
and 148, the low-$n$--high-$l$  $1i_{11/2}$ level is mostly populated. This fact is
related with the observed faster increase of the bulk contribution than of the
surface contribution in this second region.
In the third interval, with $N=148- 176$, mainly three orbitals of different
type, i.e., the $1j_{15/2}$, $2g_{7/2}$, and $3d_{5/2}$ orbitals, are populated
between $N=148$ and 176. Their
impact on the neutron skin properties is mixed and the slopes of
$\Delta r_{np}^\mathrm{bulk}$ and $\Delta r_{np}^\mathrm{surf}$  are similar
again. Finally, for $N\ge 176$, the surface part grows faster while the bulk part
slightly decreases. In this region,
two levels of the high-$n$--low-$l$ type ($4s_{1/2}$ and $3d_{3/2}$)
rapidly increase their occupancy. Their influence is magnified by
their relatively large sp mean square radii ($\lf<r^2\ri>_{nlj}> 60$ fm$^2$),
which can be seen in panel (c) of Fig. \ref{sp_pb}. It explains the fast
increase of the surface part of the NST in this region. In spite of the small
multiplicity of these levels (6 neutrons only), in the closed shell nucleus with
$N=184$ one finds the largest $\tau_{nlj}$ values for the $4s_{1/2}$ and
$3d_{3/2}$ orbitals (see Fig. \ref{sp_pb}d).


\section{Conclusions}


We have studied the influence of the properties of the single-particle
orbitals of neutrons filling the valence shell on the nuclear surface and
the neutron skin. The spherical density distributions obtained with the
SLy4 mean-field interaction for the elements Ca, Ni, Zr, Sn, and Pb were
examined in our investigation. 
Though the basic results emphasized here are general enough, the fine 
details may depend to some extent on the nuclear interaction.

The single-particle mean square radii of neutron valence-shell orbitals are
typically larger than the total neutron mean square radius of the nucleus.
These neutron orbitals impact on the nuclear surface mainly by shifting or
modifying the fall-off of the neutron density distribution. Hence, the
single-particle shell structure induces changes in the behavior of the
neutron skin thickness with respect to the smooth trend of the neutron skin
described by models of average nuclear properties such as the droplet model.

The splitting of the neutron skin thickness into bulk and
surface contributions is useful because it allows one to describe in a
relatively simple way how the neutron skin is formed.
When levels with low principal quantum number $n$ and large angular momentum $l$
are populated, the bulk part of the neutron skin thickness tends to grow fast.
The neutrons from such orbitals are mainly localized at the surface of the
neutron density. They basically shift the position of the neutron surface
outwards without altering the density slope.

The single-particle density distribution of neutrons from orbitals with high
$n$ and low $l$ has a bump beyond the neutron half-density radius of the
nucleus. It enhances the diffuseness of the nuclear surface and the surface
contribution to the neutron skin thickness. These levels play also a dominant
role in the tail of the neutron density distribution at distances a few fm
outside the surface. They govern the slope of the exponential fall-off of the
neutron density in this region.

Neutron orbitals such as $3p$ in the valence shell of Sn or $4s$ in the valence
shell of Pb, have a very large single-particle mean square radius compared with
the neighboring neutron orbitals. It magnifies the role of these levels in
increasing the surface contribution to the neutron skin thickness despite the
comparatively low number of neutrons that can be accomodated in them.

In this paper we have analyzed in detail the impact of the single-particle structure on the neutron skin thickness assuming spherical symmetry. However, this condition is not always met in the whole range of mass and atomic numbers considered here, as in the case of some of the Zr isotopes. It is known that nuclear deformations are another factor that can induce deviations of the neutron skin thickness from the average trend \cite{war98}. The study of the effects of deformation is left for a future work.




\section{Acknowledgements}

This work has been partially funded by the Spanish Consolider-Ingenio 2010
Programme CPAN CSD2007-00042, by Grant No.\ FIS2011-24154 from MICINN and
FEDER (Spain), and by Grant No.\ 2009SGR-1289 from Generalitat de Catalunya.



\bibliographystyle{apsrev}

\bibliography{skin_no_url}

\end{document}